\documentclass[aps, prd, amsmath, floats, floatfix, twocolumn, nofootinbib]{revtex4-1}
\pdfoutput=1
\usepackage{graphicx}
\usepackage{color}
\usepackage{latexsym}

\usepackage{amssymb}

\newcommand{\be}{\begin{eqnarray}}
\newcommand{\ee}{\end{eqnarray}}

\newcommand{\bea}{\begin{eqnarray}}
\newcommand{\eea}{\end{eqnarray}}

\begin{document}

\title{% 1 Nonlocal Quantum Gravity-Matter Theory} % coupled to Matter}
 %2 
 %Nonlocal Spacetime and Matter} %
  %3
  Nonlocal Spacetime-Matter}
% SOLUZIONE FINALE 

%\author{Fabio Briscese}\email{briscese.phys@gmail.com, briscesef@sustc.edu.com}

\author{Leonardo Modesto}
\email{lmodesto@sustech.edu.cn}

\affiliation{Department of Physics, Southern University of Science
and Technology, Shenzhen 518055, China}

\begin{abstract}
We propose a nonlocal field theory for gravity in presence of matter consistent with perturbative unitarity, quantum finiteness, and other essential classical properties that we are going to list below. First, the theory exactly reproduces the same tree-level scattering amplitudes of Einstein's gravity coupled to matter insuring no violation of macro-causality. Second, all the exact solutions of the Einstein's theory are also exact solutions of the nonlocal theory. Finally, and most importantly, the linear and nonlinear stability analysis of the exact solutions in nonlocal gravity (with or without matter) is in one to one correspondence with the same analysis in General Relativity. Therefore, all the exact solutions stable in the Einstein's theory are also stable in nonlocal gravity in presence of matter at any perturbative order.

%such solutions at all perturbative orders and we prove that they are stable whether they are stable in Einstein's general relativity. 

%We hereby construct an extensively study general class of nonlocal theories for gravity coupled to matter that are compatible with perturbative unitary and finiteness at quantum level. These theories exactly reproduce the same tree-level scattering amplitudes of Einstein's General Relativity insuring no violation of macrocausality. Moreover, all the exact solutions of the Einstein's theory are also exact solutions of nonlocal gravity in presence of matter. Finally, we study the nonlinear stability of such solutions at all perturbative orders and we prove that they are stable whether they are stable in Einstein's general relativity. 
%

\end{abstract}

\maketitle

\section{Introduction}

%The quantum gravity issue finds a simple, compact, and elegant solution when it is framed in the context of nonlocal field theory. 
%It is often said that we do not know what quantum gravity is. 
%We often hear that we do not know what quantum gravity is. 
%Well, this statement is completely incorrect. 
%Such statement is very incorrect because 
Nonlocal field theory aims to provide a simple, compact, and elegant solution to the quantum gravity issue.  
Indeed, gravity at quantum level is not special but exactly like all the other fundamental interactions.  
%quantum electrodynamics, quantum cromodynamics, or the gauge theory describing the weak interactions. 
However, if classical gravity is described by the Einstein-Hilbert action, then the outcome of the quantization procedure shows divergences that drastically change the structure of the theory. In the technical jargon of quantum field theory, it is said that the Einstein's theory of gravity is non-renormalizable. However, there is not any inconsistency between gravity and quantum mechanics: again, gravity is just non-renormalizable and need for a completion at high energy (a new action principle). This is something that does not happen in QED and QCD, but that physicists were called to face for the case of the Fermi theory of weak interactions. The renormalizability problem of the latter theory was overcomed replacing the Fermi's Lagrangian with a non abelian gauge theory. 
Therefore, exactly like for the weak interactions, also for gravity we need a new gravitational theory able to tame the infinities.
Now we know that the only possible extension of the Eisntein-Hilbert theory, in the field theory framework, that is consistent with unitarity and renormalizability (actually finiteness to all perturbative orders in the loop expansion) is the {\em weakly nonlocal gravity}. Another possibility is provided by the Lee-Wick quantum gravity that, however, needs further prescriptions at classical and quantum level, which are not specified by the action (references will be quoted shortly).

Records show that a nonlocal gravitational theory for gravity was proposed by Krasnikov \cite{Krasnikov} and studied two years later by Kuz'min \cite{kuzmin}. However, only recently a multidimensional generalization of the theory \cite{modesto}, with particular attention to odd dimension \cite{review}, and an extension of the theory in even dimension \cite{modestoLeslaw}, has been shown to be finite at quantum level. 
Furthermore, the Cutkosky rules \cite{cutkosky} for a general nonlocal field theory has been derived in \cite{PiusSen, Briscese:2018oyx, Briscese:2021mob}, where the perturbative unitarity was proven at any order in the loop expansion including the analysis of the anomalous thresholds. The macrocausality is also secured has proven in \cite{causality,scattering} on the base of the Shapiro's time delay. 
On the other hand, local Lee-Wick quantum gravity has been proposed in \cite{Modesto:2015ozb, Modesto:2016ofr}, on the footprint of the seminal paper \cite{shapiro3}, in order to address the unitarity issue that plagues local higher derivative theories.

Despite the very encouraging, and we would like to say surprising results listed above, not much has been done for gravity in the presence of matter \cite{Universally, FiniteGaugeTheory}. A simple way to couple nonlocal gravity to matter is introducing supersymmetry. This is a relatively easy task when we have at our disposal a superspace formalism \cite{Giaccari:2016kzy}, but the construction of other theories is still incomplete, see for example the eleven dimensional supergravity \cite{Calcagni:2014vxa}. 

%In this paper we provide a recipe to construct a quite general nonlocal field theory for gravity coupled to matter %(NLGM) consistently with the linear and nonlinear stability of exact classical solutions. 
%%The recipe is based on 
%on the base of the following four main requirements: 
In this paper we provide a recipe to construct a general nonlocal field theory for gravity coupled to matter (NLGM) on the base of the following four requirements (by Einstein's theory we will mean Einstein's gravity in presence of matter):  

\begin{enumerate}
\renewcommand{\theenumi}{\roman{enumi}}
\renewcommand{\labelenumi}{\theenumi}
\item all the solutions of Einstein's gravity must be solutions of NLGM (this is an empirical requirement), 
%(ii) 
\item all the tree-level scattering amplitudes of NLGM theory must coincide with those of Einstein's theory (this requirement guarantees macro-causality),  
%(iii) 
\item the stability analysis of the exact solutions in NLGM has to be in one to one correspondence with the same analysis in Einstein's theory (namely if a solution is stable in Einstein's gravity it is stable in NLGM too), 
%(iv) 
\item the theory has to be super-renormalizable or finite at quantum level and unitary at any perturbative order in the loop expansion. 
\end{enumerate}

As a final remark, we want to emphasize that the recipe provided in this article will allow us to construct the ultraviolet completion of any local two-derivatives theory, in any dimension, and regardless of the presence of gravity.

%\vspace{-0.35cm}

%\begin{widetext}
\section{Nonlocal gravity-matter theory}% couple to matter}
In this section we first display the general theory accomplish for the requirements (i)-(iv) listed above in the previous section, and, afterwards, we will comments on such properties. Hence, let us start with the action, %\footnote{
%We remind that in curved spacetime:
%\be
%\frac{\delta \Phi_i(x)}{\delta \Phi_j(y)} = \frac{\delta^D(x-y)}{\sqrt{-g(y)}} \delta_{ij} \, . 
%\ee
%}, 
\be
&& S[\Phi_i] = \int {\rm d}^D x \sqrt{-g} \left( {\mathcal L}_\ell + E_i \, F(\Delta)_{ij} \, E_j \right) \, , 
\label{action} \\
&& S_{\ell} = \int {\rm d}^D x \sqrt{-g} \, {\mathcal L}_\ell \, ,\quad 
 \mathcal{L}_\ell = \frac{2}{\kappa^2} R + \mathcal{L}_m \, , 
  \label{localT}\\
&& E_i(x) =  \frac{\delta S_\ell }{\delta \Phi_i(x)} \, , 
\label{EH}
\ee
where $\Phi_i$ is a set of fields, placed in a vector of components labelled by the index $i$, that include the metric and the matter's fields. $F(\Delta)_{ij}(x,y)$ is a {symmetric (respect to the swap of the indexes $i,j$ together with the spacetime points $x,y$)} tensorial entire function whose argument is a tensorial differential operator $\Delta$ that we are going to construct consistently with the stability of the exact solutions of the local theory (namely solutions of the equations of motion (EoM) $E_i = 0$). Indeed, it straightforward to show that the requirement (i) is satisfied by explicitly computing the variation of the action (\ref{action}) (up to total derivative terms and operators quadratic in the EoM $E_i$). %\footnote{In order to avoid cumbersome formulas, we here omit the spacetime points in the functional derivatives. 
The EoM for the nonlocal action (\ref{EH}) 
(see appendix \ref{EoMD} for more details)
read:
%

%\end{widetext}

%Finally, the EoM for the nonlocal theory are:
\be
{\mathcal E}_k = E_k +  2 \Delta_{k i} %\left( \frac{ \delta E_i }{\delta \Phi_k }\right)  
F_{ij} E_j + O(E^2) = 0 \, . 
\label{NLEoMa}
\ee
Since $E_k$ are the Einstein's EoM and the EoM for the matter, the following implication applies, 
\be
E_k = 0 \quad \Longrightarrow \quad {\mathcal E}_ k = 0 \, ,
\label{EvNL}
\ee
where we introduced the Hessian operator of the local theory defined by (see appendix  \ref{EoMD})\footnote{For the reader who needs to see the explicit form of the Hessian, we refer to the paper \cite{LO}, where the components of $\Delta$ are provided for gravity coupled to a complex scalar, a dirac fermion, and an abelian gauge vector.}:
\be
\Delta_{ki}\equiv  \frac{ \delta E_i }{\delta \Phi_k} 
= \frac{ \delta^2 S_\ell }{\delta \Phi_k \delta \Phi_i } \,   .
 \label{Delta0}
\ee
The same property, namely that the action consists on the Lagrangian in (\ref{localT}) plus a second operator quadratic in $E_i$, secures that all the scattering amplitudes of the nonlocal theory in presence of matter are identical to those of Einstein's gravity coupled to matter. The latter statement is based on a simple generalization of the theorem already used in \cite{scattering, nonlocaldesitter, causality}. Therefore, also the requirement (ii) is satisfied. Notice that the reverse implication in (\ref{EvNL}) is in not true because the space of solutions of the nonlocal theory is generally larger then the one of the local Einstein's theory coupled to matter. 

Notice that the action (\ref{action}) is very general and the recipe defined by the equations (\ref{action}),(\ref{localT}), and (\ref{EH}) applies to any system in any dimension, starting from a $1+0$ system to a $D-$dimensional field theory. Therefore, (\ref{action}) is actually an ultraviolet completion of any field theory, including a point-like action, with an arbitrary number of fields in presence or absence of the gravitational interaction.

%For the sake of simplicity we here consider only one matter field, hence $\Phi_i = (g_{\mu\nu}, \phi)$. 
%We can explicitly evaluate 

In order to address the stability issue as stated in (iii) 
%
%{\color{red}we have to pick out the following form factor $F_{ij}$}
%
we have to pick out a form factor $F_{ij}$ satisfying the following equation\footnote{For dimensional reasons, the operator $\Delta$, being the argument of the entire function $H$, has to be devided 
%multiplied 
by a proper power of the mass scale $\Lambda$. 
%Therefore, the correct dimensionless definition of $\Delta_{\Lambda}$ is:
%\be
%\Delta_{\Lambda  ki} :=  \frac{ \delta^2 S_\ell }{\delta \Phi_k \delta \Phi_i }  \, \Lambda^{ [\Phi_k] + [\Phi_i] }
%= \Delta_{k i} \, \Lambda^{[ \Phi_k] + [ \Phi_i ] }, 
%\ee
%where $[ \Phi_k]$ stays for the mass dimension of the field $\Phi_k$. 
%
For example, in $D=4$, if we are differentiating the Einstein's EoM respect to the metric we should divide by 
$\Lambda^4$, if we differentiate the Einstein's EoM respect to a scalar field or the matter EoM respect to the metric in both cases we should divide by $\Lambda^3$, finally, if we differentiate the matter EoM respect to the scalar field we have to divide by $\Lambda^2$.
}, 
%{\color{red}\be
%F_{i j} \equiv \left( \frac{e^{H( \Delta_\Lambda)} - 1}{2 \Delta} \right)_{i j}
% \, .
%\label{FF}
%\ee}
%
%
\be
2 \Delta_{i k} F(\Delta)_{k j} \equiv \left(e^{H( \Delta_\Lambda)} - 1 \right)_{i j}
 \, .
\label{FF}
\ee
where $H(\Delta)$ is an entire analytic function (see the appendix \ref{EoMD} for the explicit construction of $F(\Delta)$). 
%where $\Delta^{\rm s }$ is the symmetric part of $\Delta$, i.e. 
%\be 
%\Delta^{\rm s }_{ij} = \frac{1}{2} \left( \Delta_{ij} + \Delta_{ji}  \right) \, , 
%\ee
Indeed, replacing (\ref{FF}) in (\ref{NLEoMa}) the EoM turn into
\be
{\mathcal E}_k &= & 
\big( e^{H(  \Delta_{\Lambda}) } \big)_{k j } \, E_j + O(E^2) = 0 \, .
\label{LEOM}
\ee
%
%whether we are differentiating respectively the Einstein's EoM, or the matter EoM (Einstein's EoM) respect to the metric (matter fields) in (\ref{Delta}).
%
Since the function in front of the local EoM $E_j$ is invertible, we can infer that the theory is ghost-free and only the perturbative degrees of freedom of Einstein's gravity in presence of matter are allow to propagate. 
As marked by $O(E^2)$ in the EoM (\ref{LEOM}) and consistently with the previous results published in \cite{StabilityMinkAO, StabilityRicciAO}, the EoM for the perturbations of the metric and all the other fields of the theory are the same as in Einstein's local gravity with matter. Hence, the stability is guarantee at any perturbative order. Let us expand on this statement.  %{\color{red}After having fixed the all gauge symmetries,} w
We can invert the exponential factor and rewrite (\ref{LEOM}) as
\be
\mathcal{\tilde{E}}_i \equiv E_i + \big( e^{-H( \Delta_{\Lambda} ) } \big)_{i k } \left[ O(E^2) \right]_k = 0 \, .
\label{tildeEoM}
\ee
Now, given an exact background solution of the NLGM theory compatible with $E_k =0$, we can derive the EoM for the perturbations defined through an expansion of the fields, and then of the EoM, in a small dimensionless parameter $\epsilon$, namely 
\be
&&\hspace{-0.5cm} 
 \Phi_i = \sum_{n=0}^{\infty} \epsilon^n \Phi_i^{(n)} \, , \\
&& \hspace{-0.5cm} 
E_k(\Phi_i) =  \sum_{n=0}^{\infty} \epsilon^n E_k^{(n)} \, , \quad 
 \mathcal{\tilde{E}}_k(\Phi_i) =  \sum_{n=0}^{\infty} \epsilon^n \mathcal{\tilde{E}}_k^{(n)} . 
 \label{ExpEpsilon}
\ee
Assuming that the fields $\Phi_i^{(0)}$ satisfy the local background EoM, namely 
\be
E^{(0)}_k( \Phi_i^{(0)} ) =0 \,  ,
\label{background}
\ee
it is extremely simple to prove the following theorem, which is a slight generalization of the theorems proved in \cite{StabilityMinkAO, StabilityRicciAO}. 

{\bf Theorem}. In the NLGM theory, all perturbations (for gravity and matter) satisfy the same EOM of the perturbations in Einstein's gravity coupled to matter, namely 
\be
\hspace{-0.6cm}
{\mathcal{ \tilde{E}}^{(n)}_k}(\Phi_i^{(n)}) = 0 \quad \Longrightarrow \quad {E^{(n)}_k} (\Phi_i^{(n)}) = 0 \,\,\,\, \mbox{for} \,\,\,\, n>0 \, ,
\label{LeoTheorem}
\ee
where the label ``$n$" stays for the perturbative expansion of the tensors ${\mathcal{\tilde{E}}_k}$ and the EoM $E_k$ at the order ``$n$" in all the perturbations $\Phi_i^{(n)}$. 

%{\bf Proof.} 
The proof is a straightforward consequence of the EoM (\ref{tildeEoM}), which coincides with the Einstein's $E_k =0$ EoM in presence of matter up to operators $O(E^2)$, and of the invertibility of the exponential form factor. The perturbative expansion that provides the details of the proof is identical to the one for pure gravity. Therefore, we remind the reader to \cite{StabilityRicciAO}, but a short proof is given in the appendix \ref{Proof}. 
In particular, it deserves to be notice that in the perturbative expansion in $\epsilon$ of equation (\ref{LEOM}) the exponential form factor always contributes at the zero order, namely $\epsilon^0$. Moreover, for the standard model of particle physics the Hessian resulting from the local action is diagonal (or constant) at the order $\epsilon^0$ (see \cite{LO}). Hence, the inversion (\ref{tildeEoM}) is actually trivial at any perturbative order.

\section{The quantum theory}
Making use of the previous results, we show perturbative unitarity and finiteness of the  theory (\ref{action}) with form factor (\ref{FF}), and Hessian operator given in (\ref{Delta0}) and (\ref{Delta}). 

In order to show the perturbative unitarity of the NLGM theory we do not need to compute the propagator because the EoM already tell us that we have only one pole in $k^2 = - m_i^2$ for each field, exactly like in the local theory (see appendix \ref{propagators}). Therefore, the singularities of the  amplitudes are obtained from the Landau's equations as in the local case (this is due to the analytic structure of the form factor $F_{ij}$) and the Cutkosky rules are the same of the local theory, too. Hence, we can export the outcome of \cite{Briscese:2018oyx} to the general theory presented in this paper, and the unitarity is guaranteed at any perturbative order in the loop expansion. We remind that the theory has to be defined in Euclidean space and the physical amplitudes are afterwards obtained employing the analytic continuation from complex to real external energies. 
Last comment is about the non-diagonal elements of the operator $\Delta$. Indeed, one can easily see that 
such components are at least linear in the fields \cite{LO} and, therefore, can not affect the propagators around the Minkowski spacetime.

We can finally address the issue of the quantum divergences. The differential operator $\Delta$ in (\ref{Delta}) is a second order differential operator, hence, it is always possible to choose a form factor $\exp H(\Delta_\Lambda)$, asymptotically polynomial \cite{review, modesto, modestoLeslaw}, to cancel all the divergences from two loops onwards. However, in even dimension we still have one loop divergences that can likely be removed by adding other operators to the action (\ref{action}) provided they are at least cubic in the Einstein's equations of motion $E_k$. For the sake of simplicity we can consider the NLGM theory in odd dimension where we do not have one-loop divergences in dimensional regularization scheme. Therefore, the theory proposed in this paper is surely finite in odd dimensions and super-renormalizable in even dimensions.

\section{Conclusions}
We have explicitly constructed a nonlocal theory for all fundamental interactions, including gravity, that at classical level has the same solutions and the same stability properties of the local Einstein's theory in presence of matter. 
Moreover, the theory reproduces all and only the same tree-level scattering amplitudes of the local standard model of particle physics in presence of gravity, securing that there is no causality violation \cite{causality, Briscese:2019twl}. 
At quantum level, the theory is unitary and surely finite in odd dimension, while in even dimension there are only one-loop divergences that can be removed adding few more local operators on the footprint of what has been done for pure gravity without and with the cosmological constant \cite{modestoLeslaw, nonlocaldesitter}. 

Thus, this paper lays strong foundations for an ultraviolet completion of the standard model of particle physics and gravity. 

\acknowledgments
%\section{Acknowledgments} 
The author thanks F. Briscese, L. Rachwa\l, and G. Calcagni.

\begin{widetext}
\appendix 
\section{Equations of Motion} \label{EoMD}%Appendix A}
In this section we derive the EoM for a general nonlocal theory providing all the details of the calculation. We will make use of the following definition, 
\be
 \Delta_{ki}(y,x) &\equiv&  \frac{ \delta E_i(x) }{\delta \Phi_k(y)} 
= \frac{ \delta^2 S_\ell }{\delta \Phi_k(y) \delta \Phi_i(x) } 
=
 \Delta_{ k i }(x) \, \frac{\delta^D(x-y)}{\sqrt{- g(y)} }\, ,
 \label{Delta}
\ee
and we will use the functional derivative consistent with the %definition of the 
Dirac delta distribution in curved spacetime,
\be
\frac{\delta \Phi_i(x)}{\delta \Phi_j(y)} = \frac{\delta^D(x-y)}{\sqrt{-g(y)}} \delta_{ij} \,  . 
%\frac{1}{\Lambda^D} . 
\label{FuncId}
\ee
The variation of the action, introducing the short notation for the integral measure $\int d\mu(x) \equiv \int d^D x \sqrt{-g (x)}$ or simply $\int_x$, reads:
\be
\hspace{-0.7cm}
  \delta S & = & \! \int \! {\rm d} \mu(x) \, \left(\delta \Phi_i \, E_i +   \delta E_i  \, F_{ij} \, E_j + 
   E_i  \, F_{ij} \, \delta E_j 
   +
  O(E^2) 
\right) , \nonumber \\
&=& 
 \! \int \! {\rm d} \mu(x)   \, \left[ \delta \Phi_i(x) \, E_i(x) +  
 \! \int \! {\rm d} \mu(y)  \left(  \frac{\delta E_i(x)}{\delta \Phi_k(y)} \delta \Phi_k(y)   \, F_{ij}(x) \, E_j(x) + 
   E_i(x)  \, F_{ij}(x) \, \ \frac{\delta E_j (x)}{\delta \Phi_k(y)} \delta \Phi_k(y)  
   +
  O(E^2) \right) \right]  
  \nonumber \\
  &=& 
 \! \int \! {\rm d}  \mu(x) \, \left[ \delta \Phi_k (x) \, E_k(x) +  
 \! \int \! {\rm d} \mu(y) \left(  \frac{\delta E_i(x)}{\delta \Phi_k(y)} \delta \Phi_k(y)   \, F_{ij}(x) \, E_j(x) + 
   E_j(x)  \, F_{j i} (x)\,  \frac{\delta E_i (x)}{\delta \Phi_k(y)} \delta \Phi_k(y)  
   +
  O(E^2) \right) \right]  \!\! , 
  \label{step1}
%  \nonumber \\
  \ee
  where in the last term we just changed name to the indexes in order to have the same $\Delta$ operator (see (\ref{Delta})) of the last by one term. Let us now introduce the following definition,  
\be
v_i(x) =  \int \! {\rm d} ^D y \sqrt{- g(y)} \,  \Delta_{k i} (y, x)  \delta \Phi_k(y) 
\equiv  \int \! {\rm d} \mu(y) \,  \Delta_{k i} (y, x)  \delta \Phi_k(y)
\, .
\label{vu}
\ee
Therefore, replacing the definitions (\ref{Delta}) and (\ref{vu}) in (\ref{step1}) we get:
  \be
 \delta S   &=& 
 \! \int \! {\rm d} \mu(x) \, \Bigg[ \delta \Phi_k (x) \, E_k(x) +  
 \! \underbrace{ \int \! {\rm d} \mu(y) \Big( \Delta_{k i} (y, x)  \delta \Phi_k(y) }_{v_i(x)}  \, F_{ij}(x) \, E_j(x) + 
   E_j(x)  \, F_{j i} (x)\,  \Delta_{k i} (y, x) \delta \Phi_k(y)  
   +
  O(E^2) \Big) \Bigg] 
  \nonumber \\
%  &=& 
% \! \int \! {\rm d} ^D x \sqrt{- g(x)} \, \left[ \delta \Phi_k (x) \, E_k(x) +  
% \! \int \! {\rm d} ^D y \sqrt{- g(y)} \,  \Delta_{k i} (y, x)  \delta \Phi_k(y)   \Big(   
% F_{ij}(x) \, E_j(x) +   E_j(x)  \, F_{j i} (x)  \Big) 
%   +
%  O(E^2) \right]  
%   \nonumber \\
  &=& 
 \! \int \! {\rm d} \mu(x) \, \left[ \delta \Phi_k (x) \, E_k(x) +  
 v_i(x)     
 F_{ij}(x) \, E_j(x) +   E_j(x)  \, F_{j i} (x)  v_i(x)
   +
  O(E^2) \right]  
    \nonumber \\
   &=& 
 \! \int \! {\rm d} \mu(x) \,  \delta \Phi_k (x) \, E_k(x) 
 +  
 \! \int \! {\rm d} \mu(x) \, 
  \! \int \! {\rm d} \mu(y) \, \Big( 
 v_i(x)     
 F_{ij}(x,y) \, E_j(y) +   E_j(x)  \, F_{j i} (x,y)  \, v_i(y) \Big)
   +
  O(E^2) \, .
 % \nonumber \\
%   &=& 
% \! \int \! {\rm d} ^D x \sqrt{- g(x)} \,  \delta \Phi_k (x) \, E_k(x) 
% +  
% \! \int \! {\rm d} ^D x \sqrt{- g(x)} \, 
%  \! \int \! {\rm d} ^D y \sqrt{- g(y)} \, \Big( 
% v_i(x)   F_{ij}(x,y) \, E_j(y) 
% +   E_j(y)  \, F_{j i} (y,x)  \, v_i(x) \Big)
%   +
%  O(E^2) 
%  \nonumber \\
  \ee
  In the last term we now change name to the integration variables, namely $x\rightarrow y$ and $y\rightarrow x$, hence 
  \be
 \delta S     &=& 
 \! \int \! {\rm d} \mu(x) \,  \delta \Phi_k (x) \, E_k(x) 
 +  
 \! \int \! {\rm d}  \mu(x)  \, 
  \! \int \! {\rm d} \mu(y) \, \Big( 
 v_i(x)   F_{ij}(x,y) \, E_j(y) 
 +   E_j(y)  \, F_{j i} (y, x)  \, v_i(x) \Big)
   +
  O(E^2) \, . 
 % \nonumber \\
 \ee
 Making use of the following integrated symmetric property of the Hessian $\Delta$ (see the appendix (\ref{HSym})), namely
    \be
\int d^D x \sqrt{-g(x)} \int d^D y \sqrt{-g(y)} \, A_i(x) \Delta_{ij}(x,y) B_j(y) 
=
\int d^D x \sqrt{-g(x)} \int d^D y \sqrt{-g(y)} \, B_j(y) \Delta_{j i}(y,x) A_i(x) 
\, ,
\label{symH}
\ee
the variation turns into:
 \be
 \delta S      &=& 
 \! \int \! {\rm d} \mu(x) \,  \delta \Phi_k (x) \, E_k(x) 
 +  
 \! \int \! {\rm d} \mu(x) \, 
  \! \int \! {\rm d} \mu(y) \, \Big( 
 v_i(x)   F_{ij}(x,y) \, E_j(y) 
 +   v_i(x) \, F_{i j} (x,y)  \, E_j(y)  \Big)
   +
  O(E^2) 
  \nonumber \\
    &=& 
 \! \int \! {\rm d} \mu(x) \,  \delta \Phi_k (x) \, E_k(x) 
 +  
 \! \int \! {\rm d} \mu(x) \, 
  \! \int \! {\rm d} \mu(y) \, 
 2 \, v_i(x)   F_{ij}(x,y) \, E_j(y) 
   +
  O(E^2) 
  \nonumber \\
%   &=& 
% \! \int \! {\rm d} ^D x \sqrt{- g(x)} \, \left[ \delta \Phi_k (x) \, E_k(x) +  
% \! \int \! {\rm d} ^D y \sqrt{- g(y)} \,  \delta \Phi_k(y)  \, \Delta_{k i} (y, x)    \Big(   
% F_{ij}(x)  +   F_{j i} (x)  \Big) E_j(x) 
%   +
%  O(E^2) \right]  
%  \nonumber \\
%   &=& 
% \! \int \! {\rm d} ^D x \sqrt{- g(x)} \, \left[ \delta \Phi_k (x) \, E_k(x) +  
% \! \int \! {\rm d} ^D y \sqrt{- g(y)} \,  \delta \Phi_k(y)  \, \Delta_{k i} (y, x)   
%\! \int \! {\rm d} ^D x \sqrt{- g(z)} \,
%  \Big(   
% F_{ij}(x,z)  +   F_{j i} (x,z)  \Big) E_j(z) 
%   +
%  O(E^2) \right]  
%  \nonumber\\
%    &=& 
% \! \int \! {\rm d} ^D x \sqrt{- g(x)} \, \left[ \delta \Phi_k (x) \, E_k(x) +  
% \! \int \! {\rm d} ^D y \sqrt{- g(y)} \,  \delta \Phi_k(y)  \, \Delta_{k i} (y, x)   
%\! \int \! {\rm d} ^D x \sqrt{- g(z)} \,
%  \Big(   
% F_{ij}(x,z)  E_j(z) +   E_j(z) F_{j i} (z,x)  \Big)
%   +
%  O(E^2) \right]  
%   \nonumber \\
%    &=& 
% \! \int \! {\rm d} ^D x \sqrt{- g(x)} \, \left[ \delta \Phi_k (x) \, E_k(x) +  
% \! \int \! {\rm d} ^D y \sqrt{- g(y)} \,  \delta \Phi_k(y)  \, \Delta_{k i} (y, x)   
%\! \int \! {\rm d} ^D z \sqrt{- g(z)} \,
%  \Big(   
% F_{ij}(x,z)  E_j(z) +   E_j(z) F_{i j} (x,z)  \Big) 
%   +
%  O(E^2) \right]  
%   \nonumber\\
%    &=& 
% \! \int \! {\rm d} ^D x \sqrt{- g(x)} \, \left[ \delta \Phi_k (x) \, E_k(x) +  
% \! \int \! {\rm d} ^D y \sqrt{- g(y)} \,  \delta \Phi_k(y)  \, \Delta_{k i} (y, x)   
%\! \int \! {\rm d} ^D z \sqrt{- g(z)} \,
%  \Big(   
% F_{ij}(x,z)  E_j(z) +   F_{i j} (x,z)  E_j(z)  \Big)
%   +
%  O(E^2) \right]  
%  \nonumber\\
    &=& 
 \! \int \! {\rm d} \mu(x) \, \Big[ \delta \Phi_k (x) \, E_k(x) +  
\underbrace{ \! \int \! {\rm d} ^D y \sqrt{- g(y)} \,  \delta \Phi_k(y)  \, \Delta_{k i} (y, x)  }_{v_i(x) }
\! \int \! {\rm d} ^D z \sqrt{- g(z)}
 \,
 2   \, 
 F_{ij}(x,z)   E_j(z)  
   +
  O(E^2) \Big]  \, . 
\ee
%where we introduce the measure $d\mu(x) \equiv d^D x \sqrt{-g}(x)$ and 
%defined 
%\be
%v_i(x) =  \int \! {\rm d} ^D z \sqrt{- g(z)} \,  \Delta_{k i} (z, x)  \delta \Phi_k(z) \, .
%\ee
Notice that %the commutations operated 
%in the last by two step 
the operators 
\be
v_i(x)   \, , \quad  \Delta_{i j}(x,y)
 \, , \quad  F_{i j}(x,y) \, , \quad  E_j(y), 
 \label{commutativity}
\ee
%
%can be swapped 
can be freely interchanged because each of them is in a closed integral form, namely they are not differential operators acting on their right or left side but they are actually integrated quantities in which differential operators, if any, act on Dirac's delta distributions.
%
%are allowed because all the operators are independent before to integrate by parts. Indeed, all derivatives, if any, do not act on operators on the right or on the left, but on Dirac's delta distributions. 
%Moreover, it is proved in the {\color{red}appendix} that 
%\be
%\int d^D x \sqrt{-g(x)} \int d^D y \sqrt{-g(y)} \, A_i(x) \Delta_{ij}(x,y) B_j(y) 
%=
%\int d^D x \sqrt{-g(x)} \int d^D y \sqrt{-g(y)} \, B_j(y) \Delta_{j i}(y,x) A_i(x) 
%\, .
%\ee
Therefore, in a more compact and implicit notation:
\be
\delta S &=& \! \int \! {\rm d} ^D x \sqrt{- g(x)} \, \left[ \delta \Phi_k (x) \, E_k(x) +  
  \delta \Phi_k(x)  \, 2 \, \Delta_{k i} (x)    
 F_{ij}(x)   E_j(x)  
   +
  O(E^2) \right]  
  \nonumber \\
  &=& \! \int \! {\rm d} ^D x \sqrt{- g(x)} \,  \delta \Phi_k (x) \left[ E_k(x) +  
  \, 2 \, \Delta_{k i} (x)    
 F_{ij}(x)   E_j(x)  
   +
  O(E^2) \right] \, , 
\ee
and: % the EoM read:
\be 
 \mathcal{E}_l(y) &=&  \frac{\delta S}{\delta \Phi_l(y)}  =  \int \! {\rm d} ^D x \sqrt{- g(x)} \,  \frac{\delta \Phi_k(x)}{\delta \Phi_l(y)}  \,  \left[ E_k(x) +  
  \, 2 \, \Delta_{k i} (x)    
 F_{ij}(x)   E_j(x)  
   +
  O(E^2) \right] 
  \nonumber \\
  &=& 
  E_l (x) +  
  \, 2 \, \Delta_{l i} (x)    
F_{ij}(x)   E_j(x)  
   +
  O(E^2) \, ,
\ee
where we used (\ref{FuncId}). Finally, the EoM for the nonlocal theory read:
\be
{\mathcal E}_k = E_k +  2 \Delta_{k i} %\left( \frac{ \delta E_i }{\delta \Phi_k }\right)  
F_{ij} E_j + O(E^2) = 0 \, , 
\label{NLEoM}
\ee
and, making again explicit the dependence on spacetime points, the EoM (\ref{NLEoM}) should be written 
\be
 {\mathcal E}_k(x) = E_k(x) +   \int d \mu(y) \int d \mu(z) \, 2 \, \Delta_{k i }(x,y)   F_{ij}(y,z) E_j(z) + O(E^2) = 0 \, .
 \label{step2}
   \ee

In order to further simplify the above equation of motion (\ref{step2}) and rid out of the instabilities, we now expand on the operator $F_{ij}$ defined in (\ref{FF}). 
The analytic form factor $F_{ij}$ is given as a solution of the following equation,
\be
2 \Delta_{ik} F_{k j}(\Delta) = 2 F_{i k }(\Delta) \Delta_{k j} = (e^H)_{i j} - 1_{ij}.
\label{FE}
\ee
Indeed, if we define 
\be
F_{i j}(\Delta) = \sum_{n=0}^{+\infty} a_n (\Delta^n)_{i j} \qquad {\rm and} \qquad 
\left( e^{H(\Delta)} \right)_{i j} = \sum_{n=0}^{+\infty} b_n (\Delta^n)_{i j}   \, , 
\label{FeH}
\ee
by replacing (\ref{FeH}) in (\ref{FE}) we get:
\be
&& 2 \Delta \sum_{n=0}^{+\infty} a_n \Delta^n= \sum_{n=0}^{+\infty} b_n \Delta^n   -   1  \nonumber \\
&& 2 \Delta \sum_{n=0}^{+\infty} a_n \Delta^n= b_0 + \sum_{n=1}^{+\infty} b_n \Delta^n   -   1 \quad  \mbox{and assuming} \quad b_0 = 1 \quad {\rm or} \quad H(0) = 0 \, , \nonumber \\
&& 2 \Delta \sum_{n=0}^{+\infty} a_n \Delta^n =  \sum_{n=1}^{+\infty} b_n \Delta^n  \nonumber  \\
&& 2 \Delta \sum_{n=0}^{+\infty} a_n \Delta^n =  \Delta \sum_{n=1}^{+\infty} b_n \Delta^{n-1} 
\quad (n-1 = k) 
\nonumber \\
&& 2  \Delta \sum_{n=0}^{+\infty} a_n \Delta^n =  \Delta \sum_{k=0}^{+\infty} b_{k+1} \Delta^{k} \, .
\label{ExpKF}
\ee
Between the third last and the last but one step (\ref{ExpKF}) we do not need to define the inverse of $\Delta$ because in the sum on the right side we have $n>0$. Therefore, comparing the left and right side of the last equality in (\ref{ExpKF}) we figure out the relation between the coefficients $a+n$ and $b_n$, namely 
%the analytic form factor $F_{ij}$ is defined by the following coefficients, 
\be
a_n = \frac{b_{n+1}}{2} \, . 
\label{ab}
\ee
Replacing (\ref{FE}) in (\ref{step2}), 
  \be
&& 
 {\mathcal E}_k(x) = E_k(x) +  \int d \mu(z) \, 
  \left( e^{H(\Delta)} - 1 \right)_{k j}\!\! (x,z) E_j(z) + O(E^2) = 0   
  \label{EoM12}  
    \, .
\ee
where the functional identity in (\ref{EoM12}) is defined by
\be
1_{k j}(x,z) = \delta_{k j} \frac{\delta(x - z)}{\sqrt{- g(z)} } \, .
\ee
Therefore, (\ref{EoM12}) turns into: 
\be
  && 
 {\mathcal E}_k(x) = E_k(x) +    \int d \mu(z) \,  
  \left[ \left( e^{H(\Delta)} \right)_{ k j}(x,z) - \delta_{k j} \frac{\delta(x- z)}{\sqrt{- g(z)} } \right] E_j(z) + O(E^2) = 0   
  \nonumber \\
  && 
 {\mathcal E}_k(x) = E_k(x) +    \int d \mu(z) \,  
\left( e^{H(\Delta)} \right)_{ k j}(x,z) E_j(z)  -  \int d \mu(z) \delta_{k j} \frac{\delta(x - z)}{\sqrt{- g(z)} } E_j(z) + O(E^2) = 0
 \nonumber \\
  && 
 {\mathcal E}_k(x) = E_k(x) +    \int d \mu(z) \,  
 \left( e^{H(\Delta)}\right)_{ k j}(x,z) E_j(z)  - E_k(x) + O(E^2) = 0   
 \nonumber \\
  && 
 {\mathcal E}_k(x) =    \int d \mu(z) \,  
 \left( e^{H(\Delta)}\right)_{ k j }(x,z) E_j(z) + O(E^2) = 0 .
% \nonumber \\
  \ee
  Now using the second identity in (\ref{Delta}) we end up with:
  \be
&&  {\mathcal E}_k(x) =    \int d \mu(z) \,  
 \left[ \left( e^{H(\Delta)}\right)_{ k j}(x) \frac{\delta(x,z)}{\sqrt{- g(z)}} \right] E_j(z) + O(E^2) = 0
  \nonumber \\
  && 
 {\mathcal E}_k(x) =    
 \left( e^{H(\Delta_x)}\right)_{ k j} E_j(x) + O(E^2) = 0 \, .
  \label{EoM13}
\ee
The equations (\ref{FF}), (\ref{FeH}), and (\ref{ab}) allow us to avoid to invert the $\Delta$ operator. Indeed, 
it deserves to be notice that the definition of $\Delta^{-1}$ is extremely delicate. The Hessian $\Delta$ is usually not invertible because of gauge invariance\footnote{Also the operator $\Box$ is not invertible in flat space because of the zero mode/s, but not because of gauge invariance.}, and one usually adds a gauge fixing term to the (local) action in order to get the inverse. If we define 
\be
F_{i j} \equiv \left( \frac{e^{H( \Delta_\Lambda)} - 1}{2 \Delta} \right)_{i j} 
 \, , 
\label{FFb}
\ee
instead of (\ref{FF}) or (\ref{FE}), then $\Delta^{-1}$ will be part of the definition of the theory and one might worry about  an explicit break of the gauge invariance of the theory because of the gauge fixing term. However, it will not be the case because $\Delta^{-1}$ would appear in the intermedium steps of our derivation, but not in the final action and in the EoM. Indeed,  the function $F_{i j}$ is analytic in $\Delta$ and the presence of $\Delta^{-1}$ in (\ref{FFb}) is just formal. 
If we wanted to be mathematically rigorous we should add a gauge fixing term to 
%for the local action in (\ref{localT}), (\ref{EH}), and 
(\ref{Delta0}), namely 
\be
\Delta \,\, \rightarrow \,\, \Delta + H_{\rm GF} \, , 
\ee
%and accordingly in the nonlocal action (\ref{action}). So far, 
which is invertible. 
%and the Hessian $\Delta$ will turn out to be invertible. 
%
However, in the EoM and in the action $\Delta^{-1}$ would disappear because it is always multiplied by $\Delta$. Therefore, we can safely take the limit of zero gauge fixing parameters to finally recover gauge invariance for the action and the EoM.

\section{Propagators} \label{propagators}
%
%Pure gravity:
%
%
%
In order to implement the recipe developed in the main text, we here consider two explicit examples: pure gravity and a general scalar theory. In particular, we will derive the tree-level propagator in both  cases. 

\subsection{Graviton propagator} 
As a first example we consider the purely gravitational theory for which 
%, which one can get from (\ref{murho}) removing the matter's contribution. 
  the only non zero component of the $\Delta$ operator is $\Delta_{11}$, 
%\be
%\Delta_{\alpha \beta, \mu\nu}   =  \frac{1}{2} g_{\rho \sigma} E_{\mu\nu} 
% +  \frac{1}{\sqrt{-g}} \frac{\delta^2 [\sqrt{-g} \mathcal{L}_g  ] }{ \delta g^{\mu\nu} \delta g^{\alpha\beta}} \, ,\label{deltaG0} 
%\ee
but in the flat spacetime background $g_{\mu\nu} = \eta_{\mu\nu}$ and $E_{\mu\nu}=0$ (Einstein's EoM), hence 

\be
\Delta_{\mu\nu, \alpha\beta}(y, x) = \frac{\delta^2 S_\ell}{ \delta g^{\mu\nu}(y) \delta g^{\alpha \beta} (x)}
 & = &  \frac{2}{\kappa^2} \left[ 
  \frac{\delta^2 \left( \int {\rm d}^D z \sqrt{-g(z)} \mathcal{L}_g(z)  \right) }{ \delta g^{\mu\nu}(y) \delta g^{\alpha \beta} (x)}\right]_{\bf g = \eta} 
\nonumber  \\
& = &   \frac{2}{\kappa^2} \left[ 
 \frac{\delta}{\delta g^{\mu\nu}(y)} \left( \frac{\delta \left( \int {\rm d}^D z \sqrt{-g(z)} \mathcal{L}_g(z)  \right) }{  \delta g^{\alpha \beta} (x)} \right) \right]_{\bf g = \eta} 
 \nonumber \\
& = &  \frac{2}{\kappa^2} \left[ \frac{\delta  G_{\alpha \beta}(x) }{ \delta g^{\mu\nu}(y)}  
\right]_{\bf g = \eta} \nonumber \\
%& = &  \frac{2}{\kappa^2} \left[ \frac{\delta  }{ \delta g^{\mu\nu}} \left( \sqrt{-g} \, G_{\alpha \beta} + \sqrt{-g} \,  \, \frac{\delta R}{\delta g^{\alpha \beta}} \right) 
%\right]_{\bf g = \eta} \nonumber \\
%
 & = &  \frac{2}{\kappa^2}   \,  \left[ 
 \frac{\delta  R_{\alpha \beta}(x) }{\delta g^{\mu\nu}(y) } 
  - \frac{1}{2}   \, g_{\alpha \beta}(x) \, g^{\gamma \delta}(x) \,  \frac{\delta   R_{\gamma \delta}(x)  }{\delta g^{\mu\nu}(y) }  
\right]_{\bf g = \eta}
 \nonumber \\
 %
%
%& = &  \frac{2}{\kappa^2} \left[ \frac{1}{4} \left(\eta_{\alpha \mu} \eta_{\beta \nu} + \eta_{\beta \mu} \eta_{\alpha \nu} \right) \Box
%\right]
%\nonumber \\
& = &  \frac{2}{\kappa^2} \frac{1}{2}\left[ \frac{1}{2} \left(\eta_{\alpha \mu} \eta_{\beta \nu} + \eta_{\beta \mu} \eta_{\alpha \nu} \right) \Box_x
-  \eta_{\mu\nu} \eta_{\alpha \beta} \Box_x 
+ \dots \right] \, \delta^D(x-y)
\nonumber \\
& = &  \frac{2}{\kappa^2} \frac{1}{2}\left[ P^{(2)}(x) - (D-2) P^{(0)}(x) \right]_{\mu\nu, \alpha \beta}  \Box_x 
\, \delta^D(x-y)
\, , 
\nonumber \\
& \equiv & \Delta_{\mu\nu , \alpha \beta}(x) \, \delta^D(x-y)
%\frac{\delta^D(y-x)}{ \sqrt{-g(y)}}\Bigg|_{\bf g = \eta}
\, , 
\label{deltaG0b} 
\ee
where the ``dots'' stay for other second order derivative terms.
% contracted with operators to the right and to the left of $\Delta$.  
The first three steps in (\ref{deltaG0b}) are general for any background, while from the fourth step we restricted to the case of the Minkowski spacetime. 
Furthermore, $P^{(2)}, P^{(0)}$ are the spin-projectors in $D$-dimensions whose definitions read \cite{HigherDG, VN}: %anNieuwenhuizen}
\be
 && %\hspace{-1.5cm}
 P^{(2)}_{\mu \nu, \rho \sigma}(x) = \frac{1}{2} ( \theta_{\mu \rho} \theta_{\nu \sigma} +
 \theta_{\mu \sigma} \theta_{\nu \rho} ) - \frac{1}{D-1} \theta_{\mu \nu} \theta_{\rho \sigma} \, \, ,
 \nonumber\\
% \nonumber \\
% &&
% % \hspace{0.4cm}
%   P^{(1)}_{\mu \nu, \rho \sigma}(k) = \frac{1}{2} \left( \theta_{\mu \rho} \omega_{\nu \sigma} +
% \theta_{\mu \sigma} \omega_{\nu \rho}  +
% \theta_{\nu \rho} \omega_{\mu \sigma}  +
%  \theta_{\nu \sigma} \omega_{\mu \rho}  \right) \, , \nonumber   \\
   &&
  %\hspace{-1.5cm}
 P^{(0)} _{\mu\nu, \rho\sigma} (x) = \frac{1}{D-1}  \theta_{\mu \nu} \theta_{\rho \sigma}  \, , \nonumber \\
 %\,\,\,\, \,\,
% &&
% \bar{P}^{(0)} _{\mu\nu, \rho\sigma} (k) =  \omega_{\mu \nu} \, \omega_{\rho \sigma} \, \, ,  \,\,\,\,\,  \nonumber \\
%%\hspace{0.4cm}
&&
\theta_{\mu \nu} = \eta_{\mu \nu} - \frac{\partial_{\mu} \partial_{\nu}}{\Box}  \, , \,\,\,\,\,
 \omega_{\mu \nu} = \frac{\partial_{\mu} \partial_{\nu}}{\Box} \, .
 \label{proje2}
\ee
Notice that the last by one equality in (\ref{deltaG0b}) is exact because the projectors reconstruct also the terms shortly indicated with dots. 

Finally, the $\Delta$-operator for the purely gravitational theory reads as follows, 
\be
 \Delta(x,y) = \frac{1}{\kappa^2} \left[ P^{(2)}(x) - (D-2) P^{(0)}(x) \right] \Box_x \, \delta^D(y-x) \, , 
\label{deltaG}
\ee
where we did not displayed the four spacetime indexes.

In (\ref{deltaG0b}) we used the following functional derivatives \cite{scattering}, 
\be
\frac{\delta R_{\alpha \beta}(x)}{\delta g^{\mu \nu}(y)} \label{deltaG-1} 
& = & \left[ 
\frac{1}{4} \left(g_{\alpha \mu} g_{\beta \nu} + g_{\beta \mu} g_{\alpha \nu} \right) \Box + \frac{1}{2} g_{\mu\nu} 
{\nabla_\alpha \nabla_\beta }
 -  \frac{1}{2} \left( g_{\alpha \mu} \nabla_\beta \nabla_\nu + g_{\alpha \nu} \nabla_\beta \nabla_\mu \right) 
 \right]_x   \frac{ \delta^D(x-y)}{\sqrt{- g(y)}}
\, , \nonumber \\
g^{\alpha \beta}(x) \frac{\delta R_{\alpha \beta}(x)}{\delta g^{\mu \nu}(y)} \label{deltaG2} 
& = &
%\frac{1}{2} g_{\mu\nu} \Box + \frac{1}{2} g_{\mu\nu}  \Box
% -  \frac{1}{2} \left(  \nabla_\mu \nabla_\nu +  \nabla_\nu \nabla_\mu \right) \Box 
 %=
 %\nonumber \\
%& = & 
\left[  g_{\mu\nu}  \Box
 -  \frac{1}{2} \left(  \nabla_\mu \nabla_\nu +  \nabla_\nu \nabla_\mu \right)  \right]_x
  \frac{ \delta^D(x-y)}{\sqrt{- g(y)}}
 \, .
\ee

We can now expand the purely gravitational theory at the second order in the perturbation $h_{\mu\nu}$ in the Minkowski background, which satisfy the EoM $E_{\mu\nu}=0$, namely 
\be
S[g] = S[\bar{g}+h] = S[ \bar{g}] +  \frac{1}{2} \int d \mu(x_1) \, d \mu(x_2) \, h^{\mu\nu}(x_1) \,
\frac{ \delta^2 S[\bar{g}] }{\delta h^{\mu\nu}(x_1) \, \delta h^{\rho \sigma}(x_2) } \, 
h^{\rho \sigma} (x_2) + O(h^3) \, , \quad  h^{\alpha \beta} \equiv \delta g^{\alpha \beta} \, , 
\ee
where the second order term in the above expansion of the action reads:
\be
%\hspace{-0.3cm}
S^{(2)}_{\rm g} & =  & %\frac{1}{2}  
\int  \!  d^D x \left[ \sqrt{-g} \,  \frac{2}{\kappa^2} R + \sqrt{-g} \, E_{\alpha \beta} F( \Delta_{\Lambda})^{\alpha \beta, \mu \nu} E_{\mu \nu} \right]^{(2)} \nonumber \\
% & = &  \frac{1}{2} \int  \!  d^D x   \left[  \sqrt{-g} \, \mathcal{L}_{\rm g}  +  \sqrt{-g} \, E_{\alpha \beta} F(2 \Delta^{\rm s}_{\Lambda})^{\alpha \beta, \mu \nu} E_{\mu \nu} \right]^{(2)}  \nonumber \\
&= & 
\frac{1}{2}  \!  \int \!  d\mu_y \!  \int  \! d \mu_x   \, 
h^{\alpha \beta}(y) \left( \frac{\delta^2 S  }{\delta h^{\alpha \beta}(y) \delta h^{\mu\nu}(x) } \right) h^{\mu\nu}(x)
\nonumber \\
&= & 
\frac{1}{2}  \!  \int \!  d\mu_y \!  \int  \! d \mu_x   \,  
h^{\alpha \beta}(y) \left( \frac{\delta^2 S_\ell}{\delta h^{\alpha \beta}(y) \delta h^{\mu\nu}(x) } \right) 
h^{\mu\nu}(x)
%\right. 
\nonumber \\
%&+& \left. 
&&+
\frac{1}{2}  \!  \int \!  d\mu_y \!  \int  \! d \mu_x   \, 
h^{\gamma \delta}(y) \left[ \frac{\delta  }{\delta h^{\gamma \delta}(y) }  \frac{\delta  }{\delta h^{\rho \sigma}(x) } 
\left( \int d \mu_z  E_{\alpha \beta}(z) F( \Delta_{\Lambda}(z))^{\alpha \beta, \mu \nu} E_{\mu \nu}(z) \right) \right]
 h^{\rho \sigma}(x)
\nonumber \\
& = &% \frac{2}{\kappa^2}     
\frac{1}{2}  \!  \int_x \int_y % \!  d\mu_y \!  \int  \! d \mu_x   \, 
  \left[ 
h^{\alpha \beta}(y) \left( \frac{\delta^2 S_{\ell}  }{\delta h^{\alpha \beta}(y) \delta h^{\mu\nu}(x) } \right) h^{\mu\nu}(x)
%
%\right. \nonumber \\
%&& \left. 
+  2  \, h^{\gamma \delta}(y) \, 
\left( 
 \int_z %   d \mu_z  \, 
 \frac{\delta E_{\alpha \beta}(z)      }{ \delta h^{\gamma \delta}(y)} 
\,  F( \Delta_{\Lambda}(z))^{\alpha \beta, \mu \nu} \frac{\delta E_{\mu \nu}(z)}{ \delta h^{\rho \sigma}(x)}
\right) h^{\rho \sigma} (x)   \right] \, ,
\nonumber \\
& = &% \frac{2}{\kappa^2}     
\frac{1}{2}  \!  \int_x % \!  d\mu_y \!  \int  \! d \mu_x  
\int_y 
 \left[ 
h^{\alpha \beta}(y) \left( \frac{\delta^2 S_{\ell}  }{\delta h^{\alpha \beta}(y) \delta h^{\mu\nu}(x) } \right) h^{\mu\nu}(x)
%
%\right.
%\nonumber \\
%&& \left.
+   h^{\gamma \delta}(y) \, 
\left( 
\int_z %   d\mu(z)  \, 
2 \,
 \frac{\delta E_{\alpha \beta}(z)      }{ \delta h^{\gamma \delta}(y)} 
\,  F( \Delta_{\Lambda}(z))^{\alpha \beta, \mu \nu} \Delta(x, z)_{\rho \sigma, \mu\nu} 
\right) h^{\rho \sigma} (x)   \right]. 
\nonumber 
\ee
The overall $1/2$ factor is due to the functional expansion at the second order, while the multiplicative factor $2$ in the second term at the forth step % in the above evaluation 
comes from %the first variation. 
%Using now the integrated 
symmetric property of $\Delta$, namely $\Delta(x,z)_{\rho \sigma, \mu\nu} = \Delta(z,x)_{\mu\nu, \rho \sigma} $. Using again the latter property, 
\be
S^{(2)}_{\rm g}  & = &% \frac{2}{\kappa^2}     
\frac{1}{2}    \int   d\mu(y)   \int   d \mu(x)  \left[ 
h^{\alpha \beta}(y) \left( \frac{\delta^2 S_{\ell}  }{\delta h^{\alpha \beta}(y) \delta h^{\mu\nu}(x) } \right) h^{\mu\nu}(x)
\right.
\nonumber \\
&& \left.
+   h^{\gamma \delta}(y) \, 
\left( 
\int   d\mu(z)  \, 
2 \,
 \frac{\delta E_{\alpha \beta}(z)      }{ \delta h^{\gamma \delta}(y)} 
\,  F( \Delta_{\Lambda}(z))^{\alpha \beta, \mu \nu} \Delta(z, x)_{\mu\nu, \rho \sigma} 
\right) h^{\rho \sigma} (x)   \right] 
%\quad (\mbox{using:} \,\, 
%\Delta(x,z)_{\rho \sigma, \mu\nu} = \Delta(z,x)_{\mu\nu, \rho \sigma} )
\nonumber \\
& = &% \frac{2}{\kappa^2}     
\frac{1}{2}  \!  \int \!  d\mu(y) \!  \int  \! d \mu(x)   \left\{ 
h^{\alpha \beta}(y) 
\frac{\delta E_{\mu\nu}(x)}{ \delta h^{\alpha \beta}(y)} 
%\left( \frac{\delta^2 S_{\ell}  }{\delta h^{\alpha \beta}(y) \delta h^{\mu\nu}(x) } \right)
 h^{\mu\nu}(x)
%
%\right.
%\nonumber \\
%&& \left.
+   h^{\gamma \delta}(y) 
\left[ 
\int   d\mu(z)  \, 
2 \,
 \frac{\delta E_{\alpha \beta}(z)      }{ \delta h^{\gamma \delta}(y)} 
\,  F( \Delta_{\Lambda}(z))^{\alpha \beta, \mu \nu} \Delta(z, x)_{\mu\nu, \rho \sigma} 
\right] h^{\rho \sigma} (x)   \right\} .
\nonumber 
\ee
Changing the name of the indexes in the second quadratic term in $h_{\mu\nu}$, we get:
\be
S^{(2)}_{\rm g} = 
\frac{1}{2}  \!  \int \!  d\mu_y \!  \int  \! d \mu_x  \left\{ 
h^{\alpha \beta}(y) 
\frac{\delta E_{\mu\nu}(x)}{ \delta h^{\alpha \beta}(y)} 
 h^{\mu\nu}(x)
+   h^{\gamma \delta}(y) 
\left[ 
\int d \mu_w \!
\int   d\mu_z 
 \frac{\delta E_{\alpha \beta}(w)      }{ \delta h^{\gamma \delta}(y)} \, 
 \boxed{2 F(w,z)^{\alpha \beta, \mu \nu} \Delta(z, x)_{\mu\nu, \rho \sigma} }
\right] h^{\rho \sigma} (x)   \right\} .
\nonumber 
\ee
Now we replace the product in the box, i.e. $2 F(\Delta) \Delta$, with (\ref{FF}),  
\be
S^{(2)}_{\rm g} & = & 
\frac{1}{2}  \! \int_y   \int_x  \left\{ 
h^{\alpha \beta}(y) 
\frac{\delta E_{\mu\nu}(x)}{ \delta h^{\alpha \beta}(y)} 
 h^{\mu\nu}(x)
+   h^{\gamma \delta}(y) 
\left[ 
\int_w 
 \frac{\delta E_{\alpha \beta}(w)      }{ \delta h^{\gamma \delta}(y)} 
 \left( e^{H(\Delta)} - 1 \right)(w, x)^{\alpha \beta}{}_{\rho \sigma}
% \underline{2 F(w,z))^{\alpha \beta, \mu \nu} \Delta(z, x)_{\mu\nu, \rho \sigma} }
\right] h^{\rho \sigma} (x)   \right\}
\nonumber \\
& = & 
\frac{1}{2}  \!  \int_{y} \!  \int_{x}  \left\{ 
h^{\alpha \beta}(y) 
\frac{\delta E_{\mu\nu}(x)}{ \delta h^{\alpha \beta}(y)} 
 h^{\mu\nu}(x)
+   h^{\gamma \delta}(y) 
\left[ 
\int_{w} 
 \frac{\delta E_{\alpha \beta}(w)      }{ \delta h^{\gamma \delta}(y)} 
 \left( e^{H(\Delta)} (w, x)^{\alpha \beta}{}_{\rho \sigma}
 - \frac{\delta(w,x) \delta^{\alpha \beta}{}_{\rho \sigma}}{\sqrt{- g(x)}}
 \right)
% \underline{2 F(w,z))^{\alpha \beta, \mu \nu} \Delta(z, x)_{\mu\nu, \rho \sigma} }
\right] h^{\rho \sigma} (x)   \right\}
\nonumber \\
& = & 
\frac{1}{2}  \!  \int_{y}  \int_{x}  \left\{ 
h^{\alpha \beta}(y) 
\Delta(y,x)_{\alpha \beta, \mu\nu} h^{\mu\nu}(x)
+   h^{\gamma \delta}(y) 
\left[ 
\int_{w} 
\Delta(y,w)_{\gamma \delta, \alpha \beta}
 %\frac{\delta E_{\alpha \beta}(w)      }{ \delta h^{\gamma \delta}(y)} 
 \left( e^{H(\Delta)} (w, x)^{\alpha \beta}{}_{\rho \sigma}
 - \frac{\delta(w,x) \delta^{\alpha \beta}{}_{\rho \sigma}}{\sqrt{- g(x)}}
 \right)
% \underline{2 F(w,z))^{\alpha \beta, \mu \nu} \Delta(z, x)_{\mu\nu, \rho \sigma} }
\right] h^{\rho \sigma} (x)   \right\}
\nonumber \\
& = & 
\frac{1}{2}  \!  \int_{y}  \int_{x}  \int_w \left\{ 
  h^{\gamma \delta}(y) 
 \frac{\delta E_{\alpha \beta}(w)      }{ \delta h^{\gamma \delta}(y)} 
 \left( e^{H(\Delta)} (w, x)^{\alpha \beta}{}_{\rho \sigma}
 \right)
 h^{\rho \sigma} (x)   \right\}
\nonumber \\
& = &   
\frac{1}{2}  \int   d^D x \left[ 
h^{\alpha \beta}  \, \left( \frac{\delta E_{\gamma \delta}  }{ \delta h^{\alpha \beta}} \right) 
   \left[  e^{H( \Delta_{\Lambda})} \right]^{\gamma \delta}_{ \mu \nu }  \, 
 h^{\mu \nu }\right] \nonumber \\
 & = &    
\frac{1}{2} \int   d^D x \left[ 
h^{\alpha \beta}  \,  \left( \frac{\delta^2 S_\ell  }{\delta h^{\alpha \beta} \delta h^{\gamma \delta}} \right) 
   \left[  e^{H( \Delta_{\Lambda})} \right]^{\gamma \delta}_{ \mu \nu }  \, 
 h^{\mu \nu }\right] \nonumber \\
 & = &    
 \frac{1}{2 \kappa^2} \int   d^D x \left\{ 
h^{\alpha \beta}  \,  \left[    \left( P^{(2)} - (D-2) P^{(0)} \right) \Box 
\right]_{\alpha\beta, \gamma \delta}
   \left[  e^{H( \Delta_{\Lambda})} \right]^{\gamma \delta}_{ \mu \nu }  \, 
 h^{\mu \nu } 
 \right\} \, .
 %\nonumber \\
%
\label{Actiongravity2b}
\ee

In the last three steps we replaced $\sqrt{-g}=1$ because we are expanding around the Minkowski background. 
%at the second order in the perturbative expansion. 
In the last step of (\ref{Actiongravity2b}) we used the result (\ref{deltaG0b}). 
Finally, replacing the argument of the form factor $\exp H(\Delta_\Lambda)$ in (\ref{Actiongravity2b}) with 
(\ref{deltaG}), we get
\be
S^{(2)}_{\rm g} & = & 
  \frac{1}{2 \kappa^2} \int   d^D x \left\{ 
h^{\alpha \beta}  \,  \left[    \left( P^{(2)} - (D-2) P^{(0)} \right) \Box 
\right]_{\alpha\beta, \gamma \delta}
   \left[  e^{H \left(  \frac{1}{\kappa^2} \frac{P^{(2)} - (D-2) P^{(0)}  }{\Lambda^4} \Box  \right)} \right]^{\gamma \delta}{}_{ \mu \nu }  \, 
 h^{\mu \nu }
 \right\} \nonumber \\
&=& 
\frac{1}{2 \kappa^2} \int   d^D x \left\{ 
h^{\alpha \beta}  \,  \left[    \left( P^{(2)}  e^{H \left(  \frac{ \Box}{\kappa^2 \Lambda^4}   \right)}
- (D-2) P^{(0)}  e^{H \left(  \frac{ -  (D-2)  \Box }{\kappa^2 \Lambda^4}   \right)}
\right) \Box 
\right]_{\alpha\beta, \mu\nu}
     \, 
 h^{\mu \nu }
 \right\} \nonumber \\
 & \equiv & 
\frac{1}{2} \int   d^D x 
h^{\alpha \beta}  \,  \mathcal{O}_{\alpha\beta, \mu\nu}
     \, 
 h^{\mu \nu }
, 
\label{Actiongravity2}
\ee
from which after introducing the gauge fixing it is obtained the following gauge independent part of the graviton propagator \cite{HigherDG},  
\be
\mathcal{O}^{-1} = \kappa^2 \left[ \frac{P^{(2)}}{ \Box \, e^{H \left(  \frac{\Box}{\kappa^2 \Lambda^4}   \right)}} - \frac{P^{(0)}}{\Box \, (D-2)  
e^{H \left(  \frac{ - (D-2)  \Box }{\kappa^2 \Lambda^4}   \right)}} \right] \, . 
\ee
The tree-level unitarity is guarantee whether the asymptotically polynomial entire function $H(z)$ is such that $H(z)=0$. Moreover, $H(z) = H(-z)$ in order to ensure the super-renormalizablity of the theory.  
%\be 
%&& \exp H = \sum_{n=0}^{\infty} a_n  \left(  \frac{2}{\kappa^2} \frac{P^{(2)} - (D-2) P^{(0)} \Box }{\Lambda^4}   \right)^n 
%=  P^{(2)} \sum_{n=0}^{\infty} a_n  \left(  \frac{2}{\kappa^2} \frac{\Box }{\Lambda^4}   \right)^n 
%+ P^{(0)} \sum_{n=0}^{\infty} a_n  \left(  \frac{2}{\kappa^2} \frac{  - (D-2)\Box }{\Lambda^4}   \right)^n 
%\nonumber \\
%\\
%&& 
%\exp - H = \sum_{n=0}^{\infty} a_n  \left(  \frac{2}{\kappa^2} \frac{ \alpha P^{(2)} + \beta  P^{(0)} \Box }{\Lambda^4}   \right)^n
%\ee
%
%
%

\subsection{Free and Interacting Scalar Felds}
For a free scalar field the nonlocal Lagrangian, the local EoM $E_m$, and the $\Delta_{22}$ operator are:
\be
&& \mathcal{L}_m = \frac{1}{2} \phi \Box \phi +  E_m F(\Delta_\Lambda)_{22} E_m \, ,  \nonumber \\
&& E_m = \Box \phi \, , \\
&& \Delta_{22} = \frac{\delta E_m}{\delta \phi} = \Box . 
\ee
Replacing $E_m$ and $\Delta_{22}$ in $\mathcal{L}_m$ we end up with the following nonlocal Lagrangian,
\be
 \mathcal{L}_m & = & \frac{1}{2} \phi \Box \phi +  (\Box \phi)  \frac{e^{H( \Box)} - 1}{2 \Box} (\Box \phi) %\nonumber \\
%& = & 
=
 \frac{1}{2} \phi \Box e^{H( \Box)} \phi \, .
\label{NLfreeS}
\ee
Therefore, the propagator is proportional to:
\be
\frac{e^{-H( \Box)}}{\Box} \, .
\ee

For an interacting scalar field, whose local Lagrangian and EoM read
\be
&& \mathcal{L}_m^{(\rm loc)} = \frac{1}{2} \phi \Box \phi - V(\phi) \, , \\
&& E_m = \Box \phi - V^\prime(\phi) \, ,
\ee
the nonlocal theory is:
\be
&& \mathcal{L}_m = \frac{1}{2} \phi \Box \phi - V(\phi) + \left( \Box \phi - V^\prime(\phi)  \right) F( \Delta)_{22}
\left( \Box \phi - V^\prime(\phi)  \right) 
\, ,  \nonumber \\
&& \Delta_{22} = \Box - V^{\prime \prime} (\phi) \,  , \nonumber  \\
&&  F( \Delta^{\rm s}_\Lambda)_{22} = \frac{e^{H  \left( \Box  - V^{\prime \prime}(\phi)  \right)} - 1}{2  \left( \Box  - V^{\prime \prime}(\phi)  \right)} \, .
\label{NLintS}
\ee
If we switch off the interaction, the Lagrangian (\ref{NLintS}) turns into (\ref{NLfreeS}).

\section{Symmetry properties of the Hessian} \label{HSym}
In this section we explicitly prove that the Hessian operator $\Delta$ is symmetric, namely 
  \be
\boxed{\int d x  \int d y  \, A_i(x) \Delta_{ij}(x,y) B_j(y) 
=
\int d x  \int d y  \, B_j(y) \Delta_{j i}(y,x) A_i(x) }
\, .
\ee
 In order to prove the above statement is sufficient to consider the following quite general action operator, 
\be
S = \int d x \, \Phi_1(x) \, (\partial_x^n \Phi_2(x)) \, \Phi_3(x) \, , 
\label{Proto}
\ee
where $\Phi_1(x), \Phi_2(x), \Phi_3(x)$ are three general fields. For the sake of simplicity we assume $\Phi_3(x)$ to be an external classical field and we compute the components of the Hessian's operator, which is a $2\times 2$ matrix in the space of the fields $\Phi_1(x)$ and $\Phi_2(x)$, 
\be
&& 
\frac{\delta S}{\delta \Phi_1(y) } =  \int d x \, \delta(x-y) \, (\partial_x^n \Phi_2(x)) \, \Phi_3(x) 
%= (\partial_y^n \Phi_2(y)) \, \Phi_3(y) 
\, , \\
&& 
\Delta_{2 1}(z,y) = \frac{\delta^2 S}{\delta \Phi_2(z) \delta \Phi_1(y) } =  
(-1)^n \, \int d x \,  \delta(x-z) \partial_x^n \Big( \delta(x-y) \Phi_3(x) \Big) 
=  
(-1)^n  \partial_z^n \Big( \delta(z-y) \Phi_3(z) \Big) 
\label{D21}
\, , \\
&& 
\frac{\delta S}{\delta \Phi_2(y) } =  (-1)^n \int d x \,  \delta(x-y) \, \partial_x^n \left( \Phi_1(x)  \, \Phi_3(x) \right)
\, , \\
&& 
\Delta_{1 2}(z,y) = \frac{\delta^2 S}{\delta \Phi_1(z) \delta \Phi_2(y) } =  
(-1)^n  (-1)^n \int d x \,  \delta(x-z)\, \Phi_3(x) \, \partial_x^n \left(  \delta(x-y)  \right) 
=   \Phi_3(z) \, 
 \partial_z^n \Big( \delta(z-y) \Big) 
 \label{D12}
\, .
\ee
In deriving the components of the Hessian we integrating by parts several times. Moreover, the Hessian operator has zero diagonal elements because the action (\ref{Proto}) is linear in all fields.

Given two general fields $A(z)$ and $B(y)$, we now prove the following identity,
\be
\int d z \int d y A(z) \Delta_{2 1}(z , y) B(y) = \int d z \int d y \, B(y) \Delta_{1 2}(y,z) A(z) \, .
\label{ADB}
\ee
Replacing (\ref{D21}) in the left hand side of (\ref{ADB}) we find:
\be
\int d z \int d y \,  A(z) \Delta_{2 1}(z , y) B(y) & = &
\int d z \int d y \,  A(z)  \left[ (-1)^n  \partial_z^n \Big( \delta(z-y) \Phi_3(z) \Big) \right]  B(y) \nonumber \\
 & = &
(-1)^n  
\int d z \int d y  \, (-1)^n \,  \left( \partial_z^n A(z) \right) \, \delta(z-y) \, \Phi_3(z) B(y) \nonumber \\
& = &  \int d y \,  \left( \partial_y^n A(y) \right)   \Phi_3(y) B(y) \nonumber \\
& = &  \int d y \,  (-1)^n \, A(y) \,  \partial_y^n \left( \Phi_3(y) B(y) \right)  \, .
\label{ADB2}
\ee
Similarly, we replace (\ref{D12}) in the right hand side of (\ref{ADB}), 
\be
\int d z \int d y \,  B(y) \Delta_{1 2}(y , z) A(z) 
& = &\label{swap}
\int d z \int d y \,  B(y)  \left[ \Phi_3(y) \,  \partial_y^n \Big( \delta(y-z) \Big) \right]  A(z)  \\
 & = &
\int d z \int d y \,  (-1)^n \, \partial_y^n \left( B(y)  \Phi_3(y) \right) \,  \delta(y-z) \, A(z) 
 \nonumber \\
& = &  \int d y \,  (-1)^n \, A(y) \,  \partial_y^n \left( \Phi_3(y) B(y) \right)  = (\ref{ADB2}) 
 \, .
\label{ADB3}
\ee
Hence, we have proved (\ref{ADB}). 

We can also swap the fields $A(z)$ and $B(y)$ in (\ref{swap}) and the result does not change, 
\be
\int d z \int d y \,  A(z)  \Delta_{1 2}(y , z) B(y)
& = &
\int d z \int d y \,  A(z)   \left[ \Phi_3(y) \,  \partial_y^n \Big( \delta(y-z) \Big) \right]  B(y)  \\
 & = &
\int d z \int d y \,  (-1)^n \, \partial_y^n \left( \Phi_3(y) B(y)  \right) \,  \delta(y-z) \, A(z) 
 \nonumber \\
& = &  \int d y \,  (-1)^n \, A(y) \,  \partial_y^n \left( \Phi_3(y) B(y) \right)  = (\ref{ADB2}) 
 \, .
\label{ADB4}
\ee
which also proves the statement in the text below formula (\ref{commutativity}).

\section{Proof of the theorem in the main text}\label{Proof}
In order to prove the theorem about the linear and nonlinear stability of the nonlocal theory (\ref{action}), we have to expand perturbatively in $\epsilon$ (see (\ref{ExpEpsilon})) the EoM (\ref{LEOM}), 
\be 
{\bold{\mathcal{E}}}  = \bold{e^{H(  \Delta_{\Lambda}) } \, E} + O(\bold{E^2}) = 0 \, .
\label{LEOM2}
\ee
Since we want to study the stability of exact solutions of the Einstein's theory coupled to matter, we assume to expand around a metric consistent with ${\bf E^{(0)}} = 0$ (\ref{background}) (for the sake of simplicity we use the bold notation in place of the latin indexes for the fields). 

Hence, at the zero order in $\epsilon$, i.e. $\epsilon^0$, we have:
\be 
\bold{e^{H^{(0)}(  \Delta_{\Lambda}) } \, E^{(0)}} + O(\bold{E^{(0) 2}}) = 0 \, , 
\label{LEOMzero}
\ee
which is satisfied because by hypothesis ${\bf E^{(0)}} = 0$. 

At the first order $\epsilon^1$, we get:
\be 
\bold{e^{H^{(1)}(  \Delta_{\Lambda}) } \, E^{(0)}} + \bold{e^{H^{(0)}(  \Delta_{\Lambda}) } \, E^{(1)}} 
+ O(\bold{E^{(0)} E^{(1)}}) = 0 \quad \Longrightarrow \quad  {\bf E^{(1)}} = 0 \, , 
\label{LEOMuno}
\ee
where we used ${\bf E^{(0)}} = 0$. 

At the second order $\epsilon^2$, we get:
\be 
\bold{e^{H^{(2)}(  \Delta_{\Lambda}) } \, E^{(0)}} + \bold{e^{H^{(1)}(  \Delta_{\Lambda}) } \, E^{(1)}} 
+ \bold{e^{H^{(0)}(  \Delta_{\Lambda}) } \, E^{(2)}} 
+ O(\bold{E^{(1)} E^{(1)}}) 
+ O(\bold{E^{(2)} E^{(0)}}) 
= 0 \quad \Longrightarrow \quad  {\bf E^{(2)}} = 0 \, , 
\label{LEOMdue}
\ee
where we used ${\bf E^{(0)}} = 0$ and ${\bf E^{(1)}} = 0$. 

Finally, at the order $\epsilon^n$,  
\be 
&& \bold{e^{H^{(n)}(  \Delta_{\Lambda}) } \, E^{(0)}}
+ \bold{e^{H^{(n-1)}(  \Delta_{\Lambda}) } \, E^{(1)}} 
+ \bold{e^{H^{(n-2)}(  \Delta_{\Lambda}) } \, E^{(2)}} + \dots 
+ \bold{e^{H^{(0)}(  \Delta_{\Lambda}) } \, E^{(n)}} + \nonumber \\ 
&& + O(\bold{E^{(n)} E^{(0)}}) 
+ O(\bold{E^{(n-1)} E^{(1)}}) 
+ \dots
+ O(\bold{E^{(1)} E^{(n-1)}}) 
+ O(\bold{E^{(0)} E^{(n)}}) 
= 0 \quad \Longrightarrow \quad  {\bf E^{(n)}} = 0 \, , 
\label{LEOMn}
\ee
where we used: ${\bf E^{(0)}} = 0$, ${\bf E^{(1)}} = 0$, \dots, ${\bf E^{(n-1)}} = 0$. 

Therefore, 
\be
\boxed{\bold{\mathcal{E}^{(n)}} = 0 \quad \Longrightarrow \quad \bold{{E}^{(n)}} = 0} \, .
\ee

\end{widetext}


\begin{thebibliography}{99}

\bibitem{Krasnikov}
 %\cite{Krasnikov:1987yj}
%\bibitem{Krasnikov:1987yj}
  N.~V.~Krasnikov,
  ``Nonlocal Gauge Theories,''
  Theor.\ Math.\ Phys.\  {\bf 73}, 1184 (1987)
  [Teor.\ Mat.\ Fiz.\  {\bf 73}, 235 (1987)].
  %%CITATION = TMPHA,73,1184;%%
  %13 citations counted in INSPIRE as of 29 Dec 2013

\bibitem{kuzmin}
  Y.~V.~Kuz'min,
  ``The Convergent Nonlocal Gravitation. (in Russian),''
  Sov.\ J.\ Nucl.\ Phys.\  {\bf 50}, 1011 (1989)
  [Yad.\ Fiz.\  {\bf 50}, 1630 (1989)].


%\bibitem{tomboulis}
%  E.~T.~Tomboulis,
%  ``Renormalization and unitarity in higher derivative and nonlocal gravity theories,''
%  Mod.\ Phys.\ Lett.\ A {\bf 30}, 1540005 (2015). E.~T.~Tomboulis, hep-th/9702146.

\bibitem{modesto}
  L.~Modesto,
  ``Super-renormalizable Quantum Gravity,''
  Phys. \ Rev. \ D {\bf 86}, 044005 (2012)
  [arXiv:1107.2403 [hep-th]].
  
  
  %\cite{Modesto:2017sdr}
\bibitem{review} %ReModesto:2017sdr} 
  L.~Modesto and L.~Rachwal,
  %``Nonlocal quantum gravity: A review,''
  Int.\ J.\ Mod.\ Phys.\ D {\bf 26}, no. 11, 1730020 (2017).
  %doi:10.1142/S0218271817300208
  %%CITATION = doi:10.1142/S0218271817300208;%%
  %29 citations counted in INSPIRE as of 03 Apr 2019


   \bibitem{modestoLeslaw}
%\cite{Modesto:2014lga}
%\bibitem{Modesto:2014lga}
  L.~Modesto and L.~Rachwal,
  ``Super-renormalizable and finite gravitational theories,''
  Nucl.\ Phys.\ B {\bf 889}, 228 (2014)
 % doi:10.1016/j.nuclphysb.2014.10.015
  [arXiv:1407.8036 [hep-th]].
  %%CITATION = doi:10.1016/j.nuclphysb.2014.10.015;%%
  %20 citations counted in INSPIRE as of 13 Dec 2015
%\cite{Modesto:2015lna}










% UNITARITY 




\bibitem{cutkosky} R. E. Cutkosky, ``Singularities and Discontinuities of Feynman Amplitudes", Journal of Mathematical Physics {\bf 1}, 429 (1960).


%\cite{Briscese:2018oyx}
\bibitem{Briscese:2018oyx} 
  F.~Briscese and L.~Modesto,
  ``Cutkosky rules and perturbative unitarity in Euclidean nonlocal quantum field theories,''
  arXiv:1803.08827 [gr-qc].
  %%CITATION = ARXIV:1803.08827;%%
  %11 citations counted in INSPIRE as of 03 Apr 2019

%\cite{Briscese:2021mob}
\bibitem{Briscese:2021mob}
F.~Briscese and L.~Modesto,
``Non-unitarity of Minkowskian non-local quantum field theories,''
[arXiv:2103.00353 [hep-th]].
%0 citations counted in INSPIRE as of 08 Mar 2021



\bibitem{PiusSen} R. Pius, A Sen, ``Cutkosky Rules for Superstring Field Theory", JHEP {\bf 1610} (2016) 024.








% CAUSALITY 


%\cite{Giaccari:2018nzr}
\bibitem{causality} %Giaccari:2018nzr} 
  S.~Giaccari and L.~Modesto,
  ``Causality in Nonlocal Gravity,''
  arXiv:1803.08748 [hep-th].
  %%CITATION = ARXIV:1803.08748;%%
  %1 citations counted in INSPIRE as of 03 Apr 2019




%\cite{Dona:2015tra}
\bibitem{scattering} %Dona:2015tra} 
  P.~Don\'a, S.~Giaccari, L.~Modesto, L.~Rachwal and Y.~Zhu,
  ``Scattering amplitudes in super-renormalizable gravity,''
  JHEP {\bf 1508}, 038 (2015)
 % doi:10.1007/JHEP08(2015)038
  [arXiv:1506.04589 [hep-th]].
  %%CITATION = doi:10.1007/JHEP08(2015)038;%%
  %28 citations counted in INSPIRE as of 15 May 2019







% LEE-WICK QUANTUM GRAVITY 


%\cite{Modesto:2015ozb}
\bibitem{Modesto:2015ozb} 
  L.~Modesto and I.~L.~Shapiro,
  ``Superrenormalizable quantum gravity with complex ghosts,''
  Phys.\ Lett.\ B {\bf 755}, 279 (2016)
%  doi:10.1016/j.physletb.2016.02.021
  [arXiv:1512.07600 [hep-th]].
  %%CITATION = doi:10.1016/j.physletb.2016.02.021;%%
  %60 citations counted in INSPIRE as of 03 Apr 2019

%\cite{Modesto:2016ofr}
\bibitem{Modesto:2016ofr} 
  L.~Modesto,
  ``Super-renormalizable or finite Lee-Wick quantum gravity,''
  Nucl.\ Phys.\ B {\bf 909}, 584 (2016)
 % doi:10.1016/j.nuclphysb.2016.06.004
  [arXiv:1602.02421 [hep-th]].
  %%CITATION = doi:10.1016/j.nuclphysb.2016.06.004;%%
  %39 citations counted in INSPIRE as of 03 Apr 2019






% LOCAL HIGHER-DERIVATIVE QUANTUM GRAVITY 



\bibitem{shapiro3}
  M.~Asorey, J.~L.~Lopez and I.~L.~Shapiro,
  ``Some remarks on high derivative quantum gravity,''
 Int.\ J.\ Mod.\ Phys.\ A {\bf 12}, 5711 (1997)
  [hep-th/9610006].

%\bibitem{Shapirobook} I. L. Buchbinder, S. D. Odintsov, I. L. Shapiro, 
%``Effective action in quantum gravity", IOP Publishing Ltd 1992. 


% PROPAGATOR IN HIGHER-DERIVATIVE QUANTUM GRAVITY 


\bibitem{HigherDG}
%\cite{Accioly:2002tz}
%\bibitem{Accioly:2002tz}
  A.~Accioly, A.~Azeredo and H.~Mukai,
  ``Propagator, tree-level unitarity and effective nonrelativistic potential for higher-derivative gravity theories in D dimensions,''
  J.\ Math.\ Phys.\  {\bf 43}, 473 (2002).


% MATTER AND GAUGE THEORY 


%\cite{Modesto:2015lna}
\bibitem{Universally} %Modesto:2015lna} 
  L.~Modesto and L.~Rachwal,
  ``Universally finite gravitational and gauge theories,''
  Nucl.\ Phys.\ B {\bf 900}, 147 (2015)
 % doi:10.1016/j.nuclphysb.2015.09.006
  [arXiv:1503.00261 [hep-th]].
  %%CITATION = doi:10.1016/j.nuclphysb.2015.09.006;%%
  %36 citations counted in INSPIRE as of 19 Mar 2018


%\cite{Modesto:2015foa}
\bibitem{FiniteGaugeTheory} %piva} %Modesto:2015foa} 
  L.~Modesto, M.~Piva and L.~Rachwal,
  ``Finite quantum gauge theories,''
  Phys.\ Rev.\ D {\bf 94}, no. 2, 025021 (2016)
  %doi:10.1103/PhysRevD.94.025021
  [arXiv:1506.06227 [hep-th]].
  %%CITATION = doi:10.1103/PhysRevD.94.025021;%%
  %19 citations counted in INSPIRE as of 19 Mar 2018








% SUPERGRAVITY AND M-THEORY



%\cite{Giaccari:2016kzy}
\bibitem{Giaccari:2016kzy} 
  S.~Giaccari and L.~Modesto,
  ``Nonlocal supergravity,''
  Phys.\ Rev.\ D {\bf 96}, no. 6, 066021 (2017)
  %doi:10.1103/PhysRevD.96.066021
  [arXiv:1605.03906 [hep-th]].
  %%CITATION = doi:10.1103/PhysRevD.96.066021;%%
  %11 citations counted in INSPIRE as of 03 Apr 2019



%\cite{Calcagni:2014vxa}
\bibitem{Calcagni:2014vxa} 
  G.~Calcagni and L.~Modesto,
  ``Nonlocal quantum gravity and M-theory,''
  Phys.\ Rev.\ D {\bf 91}, no. 12, 124059 (2015)
  %doi:10.1103/PhysRevD.91.124059
  [arXiv:1404.2137 [hep-th]].
  %%CITATION = doi:10.1103/PhysRevD.91.124059;%%
  %56 citations counted in INSPIRE as of 03 Apr 2019


%%%% SCATTERING AMPLITUDES 



%%\cite{Dona:2015tra}
%\bibitem{scattering} %Dona:2015tra} 
%  P.~Don\'a, S.~Giaccari, L.~Modesto, L.~Rachwal and Y.~Zhu,
%  ``Scattering amplitudes in super-renormalizable gravity,''
%  JHEP {\bf 1508}, 038 (2015)
% % doi:10.1007/JHEP08(2015)038
%  [arXiv:1506.04589 [hep-th]].
%  %%CITATION = doi:10.1007/JHEP08(2015)038;%%
%  %28 citations counted in INSPIRE as of 13 Apr 2019

%\cite{Koshelev:2017ebj}
\bibitem{nonlocaldesitter} %Koshelev:2017ebj} 
  A.~S.~Koshelev, K.~Sravan Kumar, L.~Modesto and L.~Rachwal,
  ``Finite quantum gravity in dS and AdS spacetimes,''
  Phys.\ Rev.\ D {\bf 98}, no. 4, 046007 (2018)
  %doi:10.1103/PhysRevD.98.046007
  [arXiv:1710.07759 [hep-th]].
  %%CITATION = doi:10.1103/PhysRevD.98.046007;%%
  %6 citations counted in INSPIRE as of 13 Apr 2019
  
  
  
  
  
  
  %%%%%%%  STABILITY TO ALL ORDERS. %%%%%%%%
  
  
  %\cite{Briscese:2018bny}
\bibitem{StabilityMinkAO} %Briscese:2018bny} 
 %\cite{Briscese:2018bny}
%\bibitem{Briscese:2018bny} 
  F.~Briscese and L.~Modesto,
  ``Nonlinear stability of Minkowski spacetime in Nonlocal Gravity,''
  JCAP {\bf 1907}, 009 (2019)
%  doi:10.1088/1475-7516/2019/07/009
  [arXiv:1811.05117 [gr-qc]].
  %%CITATION = doi:10.1088/1475-7516/2019/07/009;%%
  %3 citations counted in INSPIRE as of 30 Nov 2019
  
  %\cite{Briscese:2019rii}
\bibitem{StabilityRicciAO} %Briscese:2019rii} 
 %\cite{Briscese:2019rii}
%\bibitem{Briscese:2019rii} 
  F.~Briscese, G.~Calcagni and L.~Modesto,
  ``Nonlinear stability in nonlocal gravity,''
  Phys.\ Rev.\ D {\bf 99}, no. 8, 084041 (2019)
%  doi:10.1103/PhysRevD.99.084041
  [arXiv:1901.03267 [gr-qc]].
  %%CITATION = doi:10.1103/PhysRevD.99.084041;%%
  %5 citations counted in INSPIRE as of 30 Nov 2019
  
  \bibitem{LO}
  %\cite{Ohta:2020bsc}
%\bibitem{Ohta:2020bsc}
N.~Ohta and L.~Rachwal,
``Effective action from the functional renormalization group,''
Eur. Phys. J. C \textbf{80}, no.9, 877 (2020)
%doi:10.1140/epjc/s10052-020-8325-8
[arXiv:2002.10839 [hep-th]].
%5 citations counted in INSPIRE as of 13 Dec 2020


  
  
  
  

% PROJECTORS

\bibitem{VN} P. Van Nieuwenhuizen, 
Nuclear Physics B 60 478-492 (1973).




% ASYMPTOTIC FREEDOM AND CAUSALITY 

%\cite{Briscese:2019twl}
\bibitem{Briscese:2019twl}
F.~Briscese and L.~Modesto,
``Unattainability of the trans-Planckian regime in nonlocal quantum gravity,''
JHEP \textbf{09}, 056 (2020)
%doi:10.1007/JHEP09(2020)056
[arXiv:1912.01878 [hep-th]].
%6 citations counted in INSPIRE as of 08 Mar 2021




\end{thebibliography}
\end{document}